# Compositionally Tunable Synthesis of High-Entropy Transition Metal Dichalcogenides Exhibiting Above-Room-Temperature Magnetism

Chen Zhang[1,†], Xiahua Zhong[1,†], Dejia Kong[2], Wenhu Wang[3], Christopher T. Nelson[2], Mingfeng Chen[4], Xiaoqing He[5], Benjamin J. Lawrie[2], Hsin-Yun Chao[2], Alex L. Melendez[2], Peixin Liu[6], Benjamin G. Gibson[1], Xiaodong Yan[6], Xi Wang[4,5], Liangbo Liang[2], Zheng Gai[2,*], Huan Zhao[2,*], and Yingchao Yang[1,*]

[1] Department of Mechanical and Aerospace Engineering, University of Missouri, Columbia, MO 65211, United States

[2] Center for Nanophase Materials Sciences, Oak Ridge National Laboratory, Oak Ridge, TN 37830, United States

[3] Frontier Institute for Research in Sensor Technologies, University of Maine, Orono, ME 04469, United States

[4] Department of Physics, Washington University in St. Louis, St. Louis, MO 63130, United States

[5] Institute of Materials Science and Engineering, Washington University in St. Louis, St. Louis, MO 63130, United States

[6] Department of Materials Sciences and Engineering, University of Arizona, Tucson, AZ 85721, United States

[*]Corresponding Authors: yingchao.yang@missouri.edu (Y.Yang), zhaoh1@ornl.gov (H.Zhao), and gaiz@ornl.gov (Z.Gai)

[†] These authors contributed equally

**Abstract**

As silicon-based electronics approach their physical limits, low-dimensional magnetic materials have attracted increasing attention for next-generation electronic and information technologies. Realizing these opportunities requires two-dimensional (2D) materials with robust and tunable room-temperature magnetism. Here, we report a two-step synthesis strategy for fabricating 2D high-entropy transition metal dichalcogenides (HE-TMDs) incorporating magnetic elements beyond conventional groups 4-6. High-entropy layered double hydroxides (HE-LDHs) are first synthesized via hydrothermal coprecipitation, enabling atomic-level mixing of diverse transition metals across groups 4-12. Subsequent chalcogenization yields 2D HE-TMDs while preserving homogeneous elemental distribution. This approach circumvents the constraints of solid-state reaction, chemical vapor transport methods, enabling the synthesis of compositionally complex and structurally controlled 2D magnets. Representative $V_\alpha Cr_\beta Fe_\gamma Co_\delta Ni_{1-\alpha-\beta-\gamma-\delta}S_{2-x}$ exhibits robust magnetism above room temperature, demonstrating the potential of this high-entropy design strategy for expanding the compositional and functional diversity of 2D magnetic materials.

## Main

Two-dimensional (2D) materials, including graphene[1,2], hexagonal boron nitride (*h*-BN)[3], black phosphorus[4], and transition-metal dichalcogenides (TMDs)[5,6], provide a wide range of basic building blocks with distinct optical and electrical properties[7]. These atomic monolayers can be further combined to create van der Waals (vdW) heterostructures, where monolayers of multiple 2D materials are either stacked vertically layer-by-layer or stitched seamlessly in-plane heterostructures[8-10]. Such vdW heterostructures have been extensively characterized and investigated for high-performance optoelectronics and electronics[11,12]. Unlike graphene, *h*-BN, and black phosphorus offer a limited number of structurally and chemically tunable derivatives, TMDs present a broader compositional space by laminating various individual transition metals between dichalcogens, allowing extensive exploration of structure-property relationships of 2D materials[13]. For example, $MoS_2$, a semiconducting layered material with a tunable bandgap of 1.2-1.8 eV, has been extensively explored for applications in nanoelectronics, particularly in field-effect transistors (FETs)[14,15]. In contrast, $VS_2$ exhibits metallic conductivity and a highly delocalized electronic structure, making it more suitable for electrochemical applications, such as supercapacitors and hydrogen evolution reaction (HER) catalysis[16,17]. The family of 2D TMDs can be further enriched by having multiple transition metal elements coexist between sulfur, selenium, tellurium or their mixtures to form binary, ternary, or quaternary compounds[18-21]. In recent work, five or more transition metals have tentatively been integrated together to realize a single-phase layered structure, 2D high-entropy TMDs (HE-TMDs), which leverage large configurational entropy to stabilize complex compositions and modulate electronic structures[22,23]. This approach shows significant potential for enhancing the catalytic activity, thermal stability, and electronic properties of TMDs[23-25].

Currently, only solid-state reaction (SSR) and chemical vapor transport (CVT) have been conducted to fabricate 2D HE-TMDs. For the SSR method, all precursors with certain stoichiometric ratio are sealed in a quartz tube under vacuum condition for chemical reaction, enabling the growth of HE-TMD crystals. For example, Ying *et al.* successfully employed the conventional SSR method to obtain various millimeter-sized polycrystalline HE-TMDs, including $(TiVCrNbTa)S_2$, $(TiVCrNbTa)Se_2$, $(TiZrHfNb)Se_2$, and $(TiVZrNbHf)Te_2$[26]. Wang *et al.* utilized the conventional SSR to synthesize HE-TMD precursors, which further undertook a molten-salt-assisted liquid-phase reaction to produce uniform and ultrathin 2D HE-TMD single crystals[22]. For the CVT approach, solid precursors typically exhibit a wide range of evaporation temperatures. To overcome this limitation, salts are commonly introduced to depress the evaporation temperatures of metals, metal oxides, or metal chlorides, enabling their simultaneous vaporization and subsequent reactions with chalcogens to form HE-TMDs[23]. For instance, Cavin *et al.* synthesized $(MoWVNbTa)S_2$, followed by a liquid-phase exfoliation to produce nanoflakes, which were further tested for $CO_2$ electroreduction. This specific catalyst exhibited exceptionally high activity and selectivity toward CO production at low overpotentials[27]. To date, the 2D HE-TMDs have exclusively incorporated transition metals from groups 4, 5, and 6, including Ti, Zr, Hf, V, Nb, Ta, Cr, Mo, and W (Fig. 1a). Such a compositional limitation is primarily governed by the Hume-Rothery criteria[28,29], which emphasize atomic-scale compatibility, including atomic size differences of less than 15%, comparable electronegativities, and the ability to form dichalcogenides with the same hexagonal close-packed (HCP) crystal structure in their individual binary phases.

Beyond the transition metals from groups 4-6, other elements, such as magnetic Fe, Co, Ni and Mn, have rarely been reported to form 2D HE-TMDs. If successfully synthesized, the combined effects of configurational entropy and lattice distortion may significantly influence super-exchange interactions by modifying intralayer cation-anion bond lengths and angles, thereby tuning the Curie temperature ($T_C$) of 2D magnetic materials[30,31]. In particular, the high-entropy effect can enhance the thermal stability of long-range magnetic ordering, as demonstrated by a recent report showing a remarkable increase in $T_C$ from 201 K to near room temperature in a $Fe_{15}Co_{15}Ni_{20}Mn_{20}Cu_{30}$ alloy[32]. Furthermore, intrinsic spin-orbit coupling can generate magnetic anisotropy, which, together with competing super-exchange pathways, may give rise to diverse magnetic ground states[33-35]. However, incorporating elements from group 7-12 is difficult because of their distinct electronic configurations, varied oxidation states, and stronger metal-metal bonding tendencies, which hinder the formation of uniform 2D HE-TMDs.

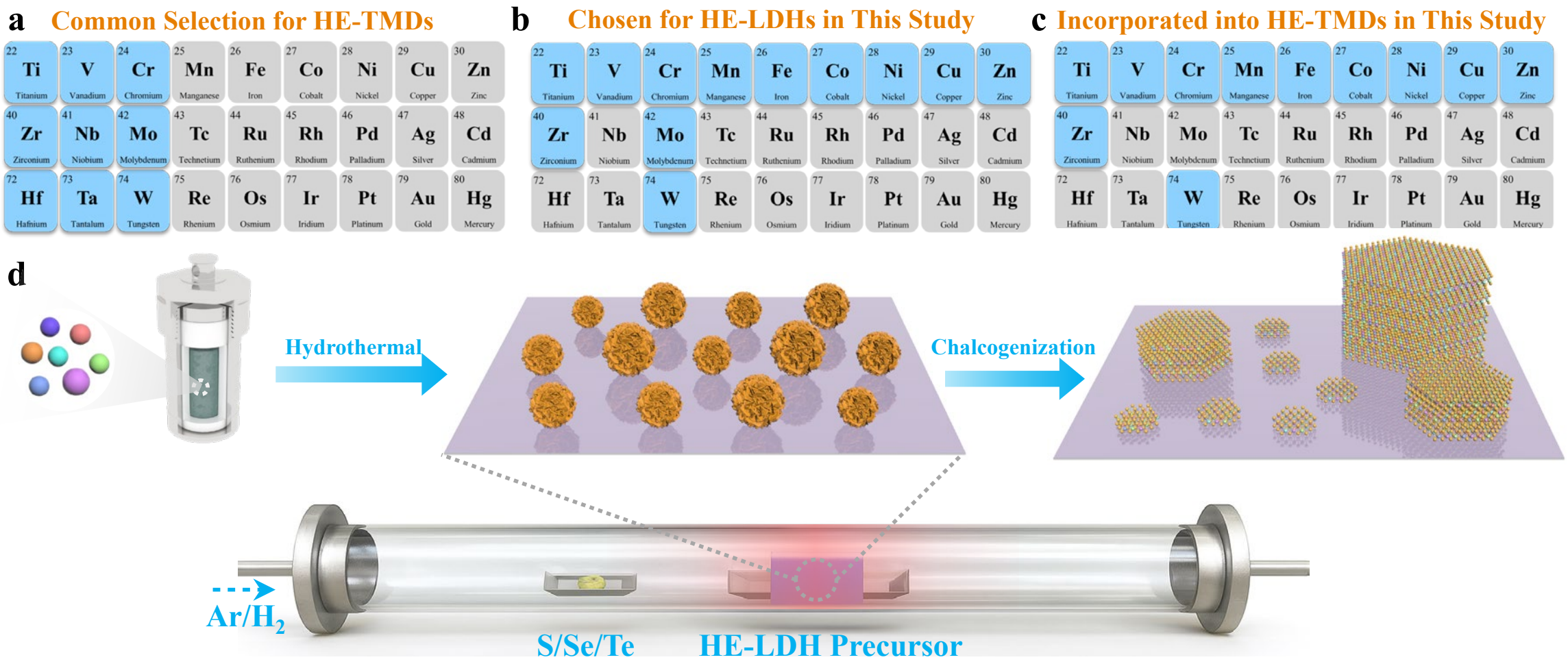

**Fig. 1| Feasible elements and synthesis strategy for the fabrication of HE-TMDs. a,** Common transition metal candidates for the synthesis of HE-TMDs using SSR or CVT. **b,** Transition metals chosen for preparing HE-LDHs. **c,** Transition metals successfully forming HE-TMDs. **d,** Schematic illustration of the two-step synthesis process.

To bypass these challenges, a facile and scalable two-step synthesis strategy was adopted to enrich the compositional space of 2D HE-TMDs, beyond the conventional element set. The first step was to employ a mild hydrothermal process to fabricate three-dimensional (3D) high-entropy layered double hydroxides (HE-LDHs) via coprecipitation[36]. Such a growth mechanism allows the coherent incorporation of multiple metals, even those with substantial differences in crystal structure, atomic radius, valence state, or electronegativity, into a single phase[37]. Therefore, the synthesis of HE-LDHs can enable the atomic-level mixing of diverse transition metals across groups 4-12 highlighted in blue in Fig. 1b and demonstrates the versatility and compositional flexibility of this class of materials. The HE-LDHs were converted into 2D HE-TMDs through a subsequent chalcogenization step, which preserved both element composition and uniform distribution inherited from the HE-LDH precursors (Fig. 1c). This two-step strategy not only circumvents the complexities of conventional methods, such as the prolonged high-temperature treatment required in SSR or the generation of intractable metal or metal oxide vapors in CVT but also offers a generalizable pathway to access compositionally complex and structurally controlled 2D HE-TMDs. In practice, 3D HE-LDHs powders were dispersed on a piece of silicon wafer coated with 300 nm $SiO_2$ and placed at the center of a tube furnace, while an alumina crucible containing sulfur, selenium, or tellurium was positioned upstream under an $Ar/H_2$ carrier gas. The chalcogen vapor reacted with the HE-LDH precursors, yielding multilayered 2D HE-TMDs, as shown in Fig. 1d.

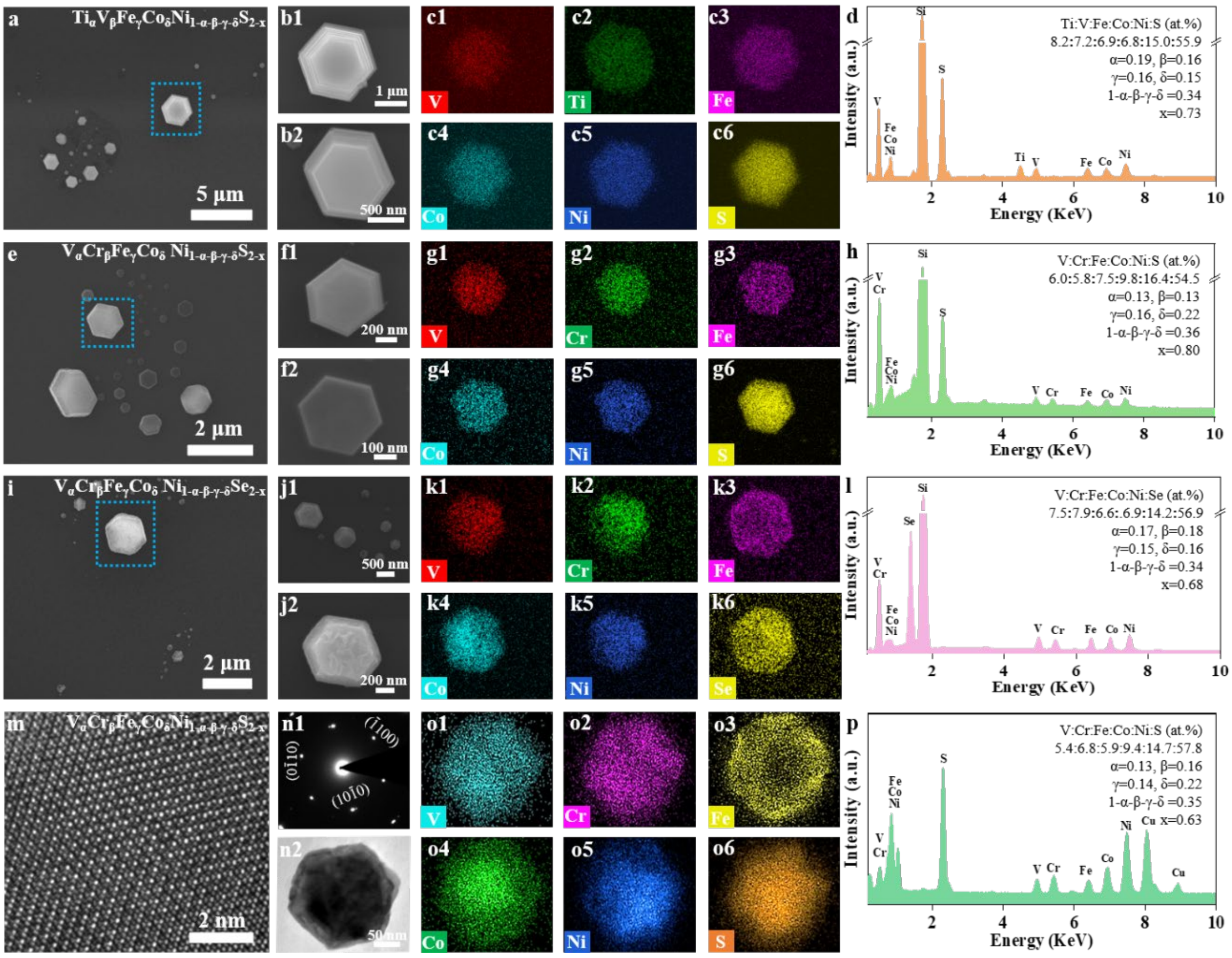

**Fig. 2| Morphology and composition of HE-TMDs. a** and **b,** SEM image of $Ti_{α}V_{β}Fe_{γ}Co_{δ}Ni_{1-α-β-γ-δ}S_{2-x}$. **c,** EDS mapping on the blue-marked region in **a**. **d,** EDS spectrum and atomic percentages of elements in $Ti_{α}V_{β}Fe_{γ}Co_{δ}Ni_{1-α-β-γ-δ}S_{2-x}$. **e** and **f,** SEM image of $V_{α}Cr_{β}Fe_{γ}Co_{δ}Ni_{1-α-β-γ-δ}S_{2-x}$. **g,** EDS mapping on the blue-marked region in **e**. **h,** EDS spectrum and atomic percentages of elements in $V_{α}Cr_{β}Fe_{γ}Co_{δ}Ni_{1-α-β-γ-δ}S_{2-x}$. **i** and **j,** SEM image of HE-TMD $V_{α}Cr_{β}Fe_{γ}Co_{δ}Ni_{1-α-β-γ-δ}Se_{2-x}$. **k,** EDS elemental mapping on the blue-marked region in **i**. **l,** EDS spectrum and atomic percentages of elements in $V_{α}Cr_{β}Fe_{γ}Co_{δ}Ni_{1-α-β-γ-δ}Se_{2-x}$. **m,** HR-TEM image of exfoliated $V_{α}Cr_{β}Fe_{γ}Co_{δ}Ni_{1-α-β-γ-δ}S_{2-x}$ crystal. **n1,** Corresponding SAED pattern obtained from the same region as m. **n2-p,** Bright-field TEM image, STEM-EDS elemental mapping, and EDS spectrum and atomic percentages of as-synthesized thin crystal.

In this study, various five-metal HE-LDH precursors were synthesized by the hydrothermal method, including ZrVFeCoNi, TiVFeCoNi, VCrFeCoNi, VWFeCoNi, VMoFeCoNi, VCoNiCuZn, and MnFeCoNiZn. The scanning electron microscope (SEM) images and corresponding energy dispersive spectroscopy (EDS) elemental mapping of all HE-LDHs are included in Supplementary Fig. 1. All specimens exhibit uniform 3D flower-like structures with most particle sizes ranging from 1 to 10 μm. The EDS mapping confirms the successful incorporation and homogeneous distribution of all constituent elements within each sample. If only the five metal elements in HE-LDHs are considered to form high-entropy alloys (HEAs), the atomic percentage of individual metals should fall into the range of 5-35 at.% according to the definition of HEAs[38,39]. The transition metal atomic percentages in HE-LDHs (Supplementary Table 1) are all within this range or a more stringent range of 15-35 at.% after normalization to HEAs (Supplementary Table 2). To understand the crystallinity of HE-LDHs, transmission electron microscopy (TEM) characterization was carried out on the selected VCrFeCoNi HE-LDH (Supplementary Fig. 2). The high-resolution TEM (HR-TEM) image and the corresponding selected-area electron diffraction (SAED) pattern reveal well-defined lattice fringes and distinct diffractions, confirming the high crystallinity of individual HE-LDH crystals. Furthermore, X-ray diffraction (XRD) analysis was performed on the selected VCrFeCoNi HE-LDH. The XRD spectrum in Supplementary Fig. 3 displays sharp diffraction peaks, implying high crystallinity of the synthesized HE-LDH across the entire sample.

With the obtained HE-LDH precursors, sulfurization, selenization, and tellurization were carried out on selected HE-LDHs. Fig. 2 shows the morphology, elemental distribution, and atomic percentages of individual elements in each HE-TMD. Figs. 2a, b present SEM images of $Ti_{\alpha}V_{\beta}Fe_{\gamma}Co_{\delta}Ni_{1-\alpha-\beta-\gamma-\delta}S_{2-x}$, where $\alpha$, $\beta$, $\gamma$, and $\delta$ are, ideally, between 0.05 and 0.35, while $x$ may vary depending on the configuration of the binary compounds. For example, if all the binary compounds have the same chemical stoichiometry as NiS, the value of $x$ will be 1, while if they all have the chemical stoichiometry of $TiS_2$, $x$ becomes 0. This definition enables quick identification of whether the obtained crystals are HE-TMDs. EDS analysis was performed on the crystal marked in Fig. 2a. The individual elemental mapping (Fig. 2c) indicates that Ti, V, Fe, Ni, Co, and S uniformly distribute throughout the crystal. The atomic ratio of Ti:V:Fe:Co:Ni:S is 0.19:0.16:0.16:0.15:0.34:1.27 (Fig. 2d), showing the metal elements fall within the defined range of HEAs. The $Ti_{\alpha}V_{\beta}Fe_{\gamma}Co_{\delta}Ni_{1-\alpha-\beta-\gamma-\delta}S_{2-x}$ crystals have well-defined HCP and layered structures with sizes ranging from hundreds of nanometers to several micrometers. An intriguing question arises as to why 3D HE-LDHs were transformed into 2D HE-TMDs with only small crystals formed on the $SiO_2$/Si substrate. Multiple processes may be the reason for such transformation. First, HE-LDHs experience dehydration around 300-700 °C, releasing $H_2O$ and forming high-entropy oxides (HEOs), as evident by thermogravimetric analysis (TGA) and XRD data shown in Supplementary Fig. 4. Furthermore, SEM images (Supplementary Fig. 5) reveal that the HEO structures remain intact when heated to 300 °C under a nitrogen atmosphere but partially collapse upon heating to 750 °C. No HCP crystalline structures are observed. Finally, when chalcogen and HEO vapors interact and nucleate on $SiO_2$/Si substrate[40], the crystal growth must accommodate multiple metal elements, which introduces greater growth resistance compared to binary TMD compounds, leading to small crystal sizes. Atomic Force Microscopy (AFM) imaging was carried out to measure the crystal thickness. Supplementary Fig. 6 shows a thickness-size dependence, where the thickness decreases from ~80 nm to ~8 nm as the lateral size decreases from 300 nm to several tens of nanometers.

Figs. 2e-l show one more high-entropy transition metal sulfide (HE-TMS), $V_{\alpha}Cr_{\beta}Fe_{\gamma}Co_{\delta}Ni_{1-\alpha-\beta-\gamma-\delta}S_{2-x}$ and one high-entropy transition metal selenide (HE-TMSe), $V_{\alpha}Cr_{\beta}Fe_{\gamma}Co_{\delta}Ni_{1-\alpha-\beta-\gamma-\delta}Se_{2-x}$. All crystals have hexagonal structures and uniform distributions of all elements. The atomic percentage of individual metal elements for both HE-TMDs almost falls within the range of 0.15 and 0.35. Comparison of $Ti_{\alpha}V_{\beta}Fe_{\gamma}Co_{\delta}Ni_{1-\alpha-\beta-\gamma-\delta}S_{2-x}$, $V_{\alpha}Cr_{\beta}Fe_{\gamma}Co_{\delta}Ni_{1-\alpha-\beta-\gamma-\delta}S_{2-x}$, and $V_{\alpha}Cr_{\beta}Fe_{\gamma}Co_{\delta}Ni_{1-\alpha-\beta-\gamma-\delta}Se_{2-x}$ indicates that the relatively higher Ni content in the HE-TMDs originates from the precursor composition used during HE-LDH synthesis. Specially, a higher Ni concentration (e.g. 1.0 mmol $Ni(NO_3)_2 \cdot 6H_2O$ versus 0.5 mmol $Co(NO_3)_2 \cdot 6H_2O$ and other transition metal precursors) was employed because it more readily promotes the formation of the LDH phase. Moreover, the uniformity of metal elements in HE-TMDs can easily be retained from the HE-LDHs. These results suggest: (1) the developed two-step synthesis approach can be extended to more 2D HE-TMDs; (2) if HE-LDHs can be synthesized, fabrication of HE-TMDs is very likely to be possible; (3) the ratio of transition metals can be carefully tuned depending on the requirement. As SEM analysis is limited to large flakes, in order to further examine if all metals exist and are uniformly distributed at nanoscale dimensions, the as-synthesized $V_{\alpha}Cr_{\beta}Fe_{\gamma}Co_{\delta}Ni_{1-\alpha-\beta-\gamma-\delta}S_{2-x}$ crystals were transferred from the $SiO_2$/Si substrate onto a copper Quantifoil TEM grid. In addition, to minimize the influence of thickness on phase uniformity and elemental distribution, crystals from the same batch were further mechanically exfoliated using adhesive tape, and the resulting thinner flakes were transferred onto TEM grids for comparison. Figs. 2m-o show the (S)TEM characterization of $V_{\alpha}Cr_{\beta}Fe_{\gamma}Co_{\delta}Ni_{1-\alpha-\beta-\gamma-\delta}S_{2-x}$ crystals. The HR-TEM image (Fig. 2m) of exfoliated crystal reveals well-resolved lattice fringes with a continuous crystalline framework, without observable domain boundaries or secondary phases, confirming the high crystallinity and single-phase nature at the nanoscale. The corresponding SAED pattern taken from the same region (Fig. 2n1) exhibits a clear hexagonal symmetry, consistent with an HCP crystal structure. High-Angle Annular Dark-Field (HAADF) STEM imaging (Supplementary Fig. 7) reveals slight variations in lattice spacing at the atomic scale, which would originate from lattice-distortion-induced displacive modulation among different transition metals[41,42]. For comparison, the SAED pattern obtained from the directly transferred thicker samples (Supplementary Fig. 8) shows satellite spots, which are primarily attributed to moiré effects arising from overlapping layers[43,44]. Bright-field TEM imaging (Fig. 2n2) on as-synthesized crystal further illustrates a well-defined hexagonal shape even when their lateral dimensions are reduced below 200 nm. Metal and sulfur elements are uniformly distributed (Fig. 2o), which is consistent with the SEM elemental mapping results. The atomic ratio of V:Cr:Fe:Co:Ni:S is 0.13:0.16:0.14:0.22:0.35:1.37 (Fig. 2p), which is very close to that obtained from SEM-EDS. The atomic percentages of metal elements are expected to vary slightly among crystals, while the overall compositional characteristics remain consistent with the definition of HEMs. Furthermore, STEM-EDS analysis of the exfoliated crystal (Supplementary Fig. 9) confirms a homogeneous distribution of all constituent elements, without evidence of local phase separation.

Beyond the two HE-TMSs and one HE-TMSe in Fig. 2, the facile two-step synthesis strategy was successfully used to fabricate four more HE-TMSs $Zr_{0.20}V_{0.08}Fe_{0.14}Co_{0.18}Ni_{0.40}S_{1.21}$, $Mn_{0.26}Fe_{0.14}Co_{0.32}Ni_{0.15}Zn_{0.13}S_{1.18}$, $V_{0.13}W_{0.11}Fe_{0.18}Co_{0.22}Ni_{0.36}S_{1.20}$, and $V_{0.13}Co_{0.10}Ni_{0.15}Cu_{0.34}Zn_{0.30}S_{1.13}$ (Supplementary Fig. 10), two more HE-TMSes $Ti_{0.16}V_{0.13}Fe_{0.21}Co_{0.20}Ni_{0.26}Se_{1.29}$ and $Zr_{0.19}V_{0.13}Fe_{0.17}Co_{0.15}Ni_{0.31}Se_{1.33}$ (Supplementary Fig. 11), as well as to tentatively synthesize $V_{\alpha}Mo_{\beta}Fe_{\gamma}Co_{\delta}Ni_{1-\alpha-\beta-\gamma-\delta}S_{2-x}$ (Supplementary Fig. 12) and three high-entropy transition metal tellurides (HE-TMTes) $Ti_{\alpha}V_{\beta}Fe_{\gamma}Co_{\delta}Ni_{1-\alpha-\beta-\gamma-\delta}Te_{2-x}$, $Zr_{\alpha}V_{\beta}Fe_{\gamma}Co_{\delta}Ni_{1-\alpha-\beta-\gamma-\delta}Te_{2-x}$, and $V_{\alpha}Cr_{\beta}Fe_{\gamma}Co_{\delta}Ni_{1-\alpha-\beta-\gamma-\delta}Te_{2-x}$ (Supplementary Fig. 13). In addition, two four-metal TMSs ($V_{\alpha}Fe_{\beta}Co_{\gamma}Ni_{1-\alpha-\beta-\gamma}S_{2-x}$ and $Cr_{\alpha}Fe_{\beta}Co_{\gamma}Ni_{1-\alpha-\beta-\gamma}S_{2-x}$) shown in Supplementary Fig. 14 have also been synthesized to compare magnetic behaviors with $V_{\alpha}Cr_{\beta}Fe_{\gamma}Co_{\delta}Ni_{1-\alpha-\beta-\gamma-\delta}S_{2-x}$. All successfully synthesized HE-TMDs exhibit well-defined hexagonal layered structures. EDS elemental mapping and spectra confirm the presence and uniform distribution of all metal elements. The atomic ratios fall within the defined HEM range. Despite multiple trials, sulfurization of the VMoFeCoNi HE-LDH did not yield the expected 2D layered structure. Instead, the product largely retained the morphology of the original HE-LDH precursor, even though all the metal elements remained uniformly distributed throughout the crystals. Moreover, the tentative tellurization of HE-LDHs delivered layered structures and homogeneous element distribution, while the tellurium content is only ~35 at.%, which is substantially lower than the sulfur and selenium concentrations typically achieved in HE-TMSs and HE-TMSes, respectively. The relatively low tellurium concentration is primarily attributed to the high susceptibility of tellurides to oxidation[45]. Through this approach, transition metals such as Ti, Zr, V, Cr, W, Mn, Fe, Co, Ni, Cu, and Zn are effectively incorporated into 2D HE-TMDs, enabling diverse elemental combinations.

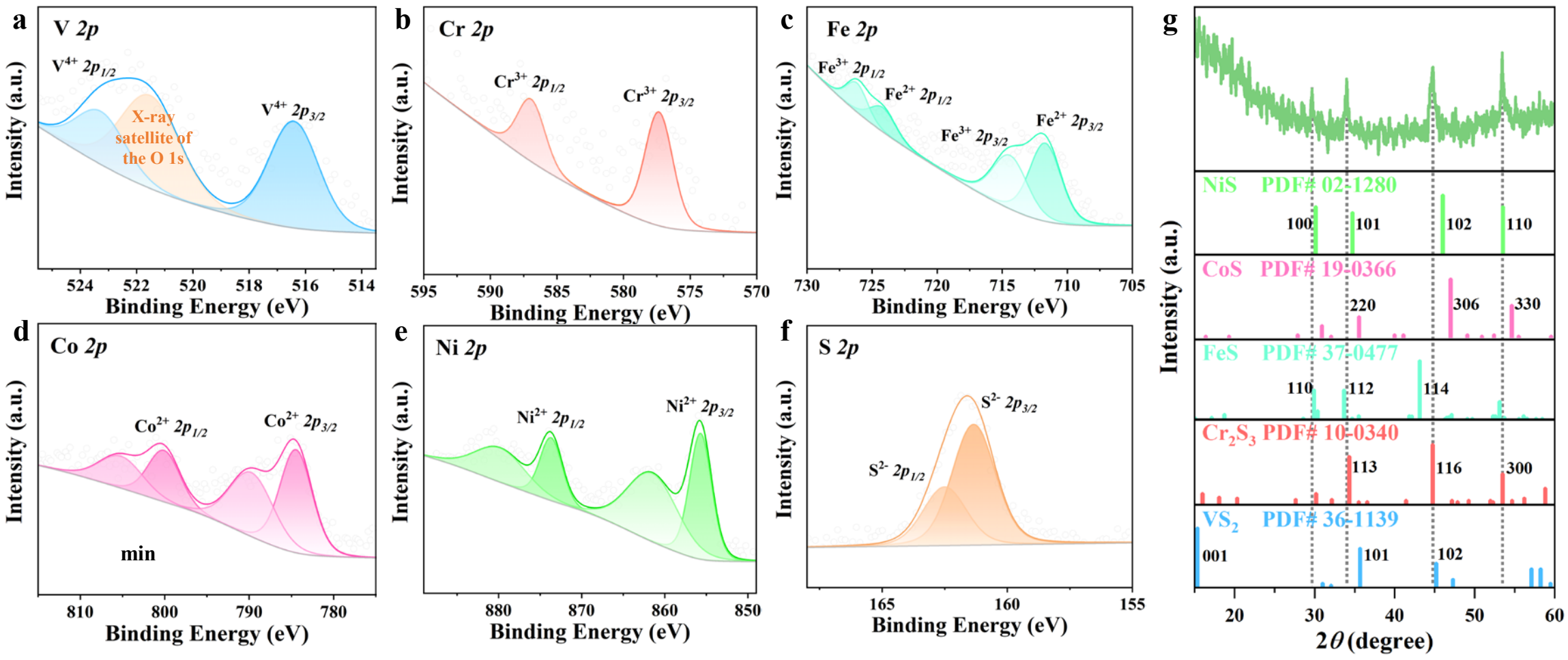

**Fig. 3| Structural and chemical characterization of $V_{\alpha}Cr_{\beta}Fe_{\gamma}Co_{\delta}Ni_{1-\alpha-\beta-\gamma-\delta}S_{2-x}$.** XPS analysis of the constituent elements: **a,** V 2*p*, **b,** Cr 2*p*, **c,** Fe 2*p*, **d,** Co 2*p*, **e,** Ni 2*p*, and **f,** S 2*p*. **g,** Comparison of XRD spectra among $V_{\alpha}Cr_{\beta}Fe_{\gamma}Co_{\delta}Ni_{1-\alpha-\beta-\gamma-\delta}S_{2-x}$, NiS, CoS, FeS, $Cr_2S_3$, and $VS_2$.

X-ray photoelectron spectroscopy (XPS) further proves the existence of all elements in $V_{\alpha}Cr_{\beta}Fe_{\gamma}Co_{\delta}Ni_{1-\alpha-\beta-\gamma-\delta}S_{2-x}$. According to the high-resolution XPS spectra of V *2p*, Cr *2p*, Fe *2p*, Co *2p*, and Ni *2p*, all metal species are in cationic form with various oxidation states, including $V^{4+}$ at 516.46 eV, $Cr^{3+}$ at 577.41 eV, $Fe^{2+}$ at 711.75 eV, $Fe^{3+}$ at 714.60 eV, $Co^{2+}$ at 784.54 eV, and $Ni^{2+}$ at 855.71 eV (Figs. 3a-e). In the XPS spectrum of $S^{2-}$ *2p*, the $S^{2-}$ $2p_{3/2}$ (161.35 eV) and $2p_{1/2}$ (162.50 eV) peaks correspond to the metal-sulfur interactions, confirming the formation of a metal sulfide (Fig. 3f). The coexistence of these diverse metal cations with oxidation states ranging from +2 to +4 leads to an average cationic valence lower than the +4 required for a stoichiometric $MS_2$ phase. To maintain global charge neutrality within this highly entropy-stabilized single-phase lattice, the anion chemistry self-regulates, resulting in a naturally stable, non-stoichiometric structure with coordination unsaturation ($S_{2-x}$)[46]. This fundamental charge-compensating mechanism aligns with the low sulfur-to-metal ratio determined by SEM and TEM EDS quantifications. Fig. 3g presents the XRD pattern of $V_\alpha Cr_\beta Fe_\gamma Co_\delta Ni_{1-\alpha-\beta-\gamma-\delta}S_{2-x}$. The four main peaks between 30° and 55° overlap or come close to those of individual binary sulfides, including NiS, CoS, FeS, $Cr_2S_3$, and $VS_2$. Such binary sulfides were chosen based on the valence states proven in XPS. The lattice spacing of $V_\alpha Cr_\beta Fe_\gamma Co_\delta Ni_{1-\alpha-\beta-\gamma-\delta}S_{2-x}$ along a specific crystallographic plane arises from the combined contributions of all participating metals. Therefore, they should differ from those along the same plane in individual binary sulfides, due to the aforementioned variations in crystal structure, atomic radius, valence state, or electronegativity among Ni, Co, Fe, Cr, and V. A similar trend is also observed in the XRD pattern of $Ti_\alpha V_\beta Fe_\gamma Co_\delta Ni_{1-\alpha-\beta-\gamma-\delta}S_{2-x}$ (Supplementary Fig. 15), reinforcing the generality of the single-phase high-entropy nature. The Raman spectrum, specifically the out-of-plane ($A_{1g}$) vibrational mode of $V_\alpha Cr_\beta Fe_\gamma Co_\delta Ni_{1-\alpha-\beta-\gamma-\delta}S_{2-x}$ (Supplementary Fig. 16), was also compared with those of reported single-metal sulfides (Supplementary Table 3). The peak at 300 $cm^{-1}$ corresponds to $Cr_2S_3$ and FeS. The peak at 344 $cm^{-1}$ is close to that of NiS. In addition, the peak located at 382 $cm^{-1}$ aligns well with those reported for CoS and $VS_2$. Such results suggest that variations in neighboring metal atoms have a limited influence on the out-of-plane lattice vibration in HE-TMDs, whereas the in-plane vibration tends to be averaged as all neighboring metal atoms coordinate together. The environmental and thermodynamic stability of the synthesized HE-TMDs was rigorously evaluated through a series of thermal treatments. Specifically, the samples were subjected to heat treatment in ambient air at 100 °C and a sulfur atmosphere at 800 °C for 2 hrs. Post-annealing characterizations via Raman spectroscopy (Supplementary Fig. 17) and XRD (Supplementary Fig. 18) reveal that the single-phase crystalline structure remained intact under both conditions.

Scanning nitrogen-vacancy (NV) microscopy (SNVM) was employed to probe the nanoscale magnetic response of the $V_{\alpha}Cr_{\beta}Fe_{\gamma}Co_{\delta}Ni_{1-\alpha-\beta-\gamma-\delta}S_{2-x}$ crystals. A diamond probe hosting a single <100> oriented NV center at the tip apex scanned across the sample surface to acquire the AFM topography and the corresponding stray field map simultaneously. A small out-of-plane magnetic bias field of 56 G was applied during all NV measurements. Fig. 4a shows the topographic image of a representative sample, consisting of flakes with lateral dimensions from ~100 nm to 1 μm and thicknesses up to 250 nm. Fig. 4b presents the corresponding stray field distribution recorded with the NV center positioned ~50 nm above the surface. Because the NV detects only the magnetic-field component projected along its quantization axis (tilted by 54.7° from the surface normal and oriented along the negative x-direction in the map), the image directly reflects that field component. Pronounced local magnetic contrasts appear at the positions of the larger flakes, unambiguously indicating that the stray field originates from the material itself, rather than the substrate. The stray field intensity exhibits a clear dependence on flake volume, while the observed magnitude and spatial variation reflect the formation of nanoscale ferromagnetic domains within larger flakes. The room-temperature reconstructed magnetization vector field (Supplementary Figs. 19-20) reveals sub-100-nm domains with comparable in-plane and out-of-plane components, indicative of mixed (easy-cone-like) anisotropy. Consequently, the magnetic easy axis orientation depends on the flake's shape and thickness due to the competing effect between magnetocrystalline and demagnetizing energies. Fig. 4c shows the stray field map of the same region measured at 410 K. Remarkably, the magnetic landscape remains largely unchanged at this temperature, demonstrating robust ferromagnetism suitable for industry or aerospace microelectronic applications which require stable operation in heated environments. This establishes $V_{\alpha}Cr_{\beta}Fe_{\gamma}Co_{\delta}Ni_{1-\alpha-\beta-\gamma-\delta}S_{2-x}$ as a 2D ferromagnet with one of the highest $T_C$ reported to date (Supplementary Table 4). Additional SNVM measurements on $V_{\alpha}Cr_{\beta}Fe_{\gamma}Co_{\delta}Ni_{1-\alpha-\beta-\gamma-\delta}S_{2-x}$ (Supplementary Fig. 21) and $Ti_{\alpha}V_{\beta}Fe_{\gamma}Co_{\delta}Ni_{1-\alpha-\beta-\gamma-\delta}S_{2-x}$ (Supplementary Fig. 22) further confirm that the spatial distribution of the stray magnetic field closely matches the sample distribution observed in the corresponding AFM images, providing evidence that the magnetic signal originates from the crystals rather than from impurities or the substrate. Moreover, thickness-dependent analysis of $Ti_{\alpha}V_{\beta}Fe_{\gamma}Co_{\delta}Ni_{1-\alpha-\beta-\gamma-\delta}S_{2-x}$ (Supplementary Fig. 23) reveals an overall increase in stray magnetic field strength with increasing flake thickness[47, 48].

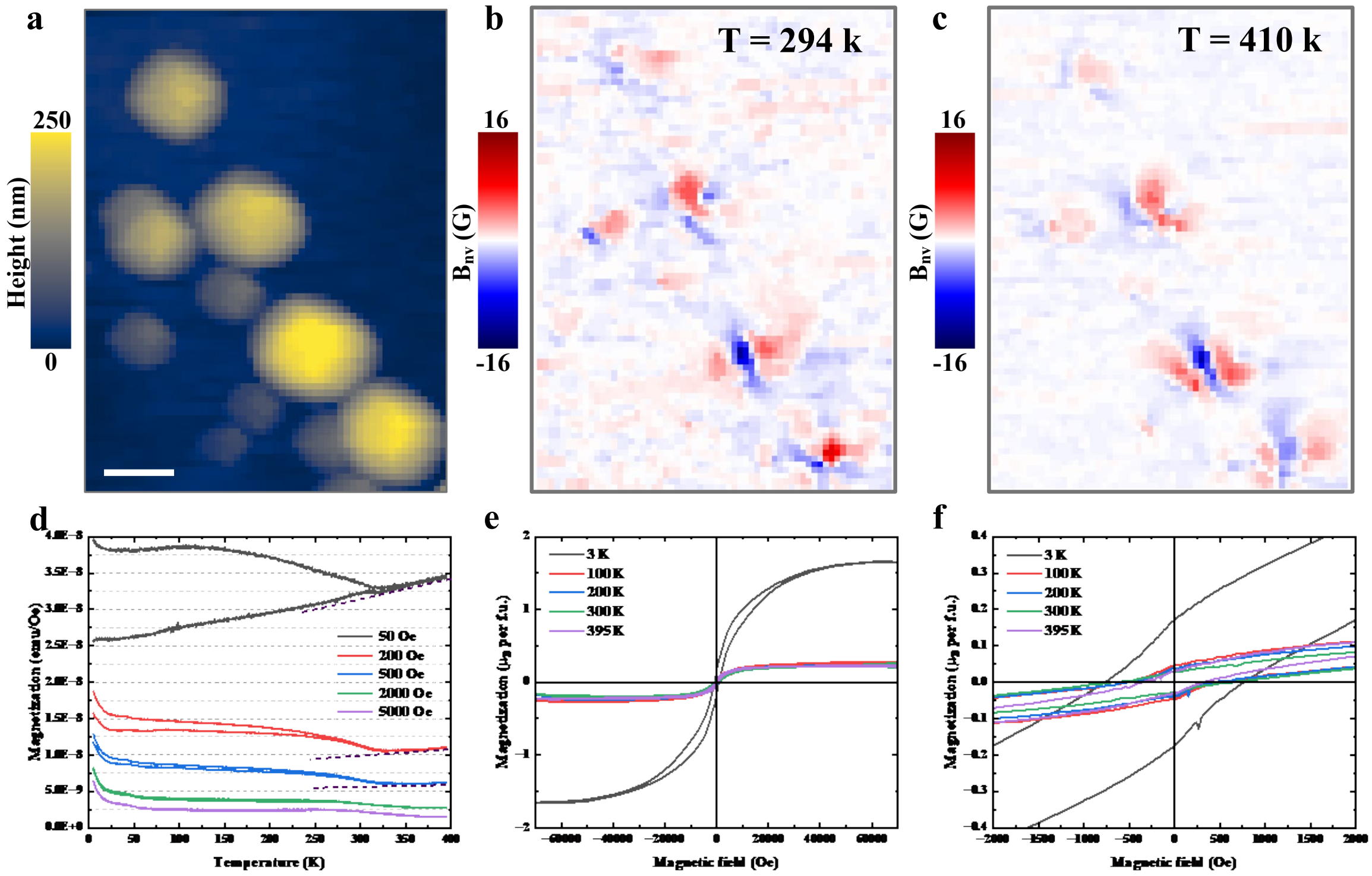

**Fig. 4| Magnetic behaviors of $V_{\alpha}Cr_{\beta}Fe_{\gamma}Co_{\delta}Ni_{1-\alpha-\beta-\gamma-\delta}S_{2-x}$ measured by SNVM and MPMS. a,** AFM topography acquired with the SNVM. **b,** Corresponding map of the room-temperature stray magnetic field projection along the NV quantization axis. **c,** The same stray magnetic field map measured at 410 K. **d,** ZFC and FC magnetization under different magnetic fields for $V_{\alpha}Cr_{\beta}Fe_{\gamma}Co_{\delta}Ni_{1-\alpha-\beta-\gamma-\delta}S_{2-x}$ crystals. A small irreversible splitting and a distinct downward kink near 320 K for lower magnetic field (50, 200, 500 Oe) is highlighted by the dashed guided lines. For easy comparison, the curves are normalized by applied fields. **e, f,** the normalized magnetic hysteresis loops for $V_{\alpha}Cr_{\beta}Fe_{\gamma}Co_{\delta}Ni_{1-\alpha-\beta-\gamma-\delta}S_{2-x}$ crystals, in the unit of $\mu_B$ per formula unit (f.u.). **e,** The whole loops at different temperatures. **f,** Low field region of the M(H) loops in **e**.

Global magnetization measured with magnetic property measurement system (MPMS) reveals that the 2D $V_{\alpha}Cr_{\beta}Fe_{\gamma}Co_{\delta}Ni_{1-\alpha-\beta-\gamma-\delta}S_{2-x}$ exhibits robust magnetism persisting beyond room temperature, confirming the NV microscopy data. Fig. 4d shows the temperature-dependent magnetization measured under zero field-cooled (ZFC) and field-cooled (FC) conditions at applied fields of 50, 200, 500, 2000, and 5000 Oe. The magnetization remains finite up to 400 K without a transition to a paramagnetic state, indicating robust magnetic ordering above room temperature. A broad anomaly is observed near 320 K. Under low applied fields (50-500 Oe), the magnetization gradually decreases upon cooling before reaching a shallow minimum near 320 K, as highlighted by the dashed guidelines. The broad anomaly near ~320 K is consistently observed across batches and measurement conditions, indicating that it reflects an intrinsic magnetic crossover. This non-monotonic behavior suggests a temperature-dependent evolution of magnetic anisotropy and competing spin alignment, where weak fields cannot fully overcome the competing anisotropy energies. As the applied field increases, this feature is progressively suppressed, indicating that stronger fields stabilize the magnetic alignment. Such systematic field dependence suggests the existence of competing spin alignments, an intrinsic spin reorientation transition associated with changes in magnetic anisotropy, rather than thermal suppression of ferromagnetism or the presence of impurity phases. The normalized magnetic hysteresis curves, expressed in $\mu_B$/f.u., are presented in Fig. 4e and f. The detailed calculation of area mass of $V_\alpha Cr_\beta Fe_\gamma Co_\delta Ni_{1-\alpha-\beta-\gamma-\delta}S_{2-x}$ crystals can be found in Supplementary Figs. 24-30 and Tables 5, 6, which was used for the normalization of magnetization data. The saturation moment reaches 1.67 ± 0.12 $\mu_B$/f.u. at 3 K. Although the saturation magnetization decreases with increasing temperature, it remains as high as 0.22 $\mu_B$/f.u. at 395 K. The remanent magnetization and coercive field are 0.17 $\mu_B$/f.u. and 720 Oe at 3 K, and 0.03 $\mu_B$/f.u. and 340 Oe at 395 K, respectively.

The NV measurements in Fig. 4a-b were performed under a small bias field (50-100 Oe), probing only a minor hysteresis loop rather than the fully saturated state measured by MPMS. Therefore, the NV-reconstructed magnetization should be compared with the near-zero-field MPMS magnetization instead of the saturation magnetization. At room temperature, MPMS gives a near-zero-field magnetization of ~0.03 $\mu_B$/f.u., while the NV reconstruction yields a local magnetization of ~0.04 $\mu_B$/f.u. (Supplementary Fig. 31), showing good agreement in order of magnitude. This consistency is reasonable because NV magnetometry probes the local magnetization of an individual representative flake, whereas MPMS measures the spatially averaged response of the entire sample, including variations in flake thickness, domain structure, and magnetic inhomogeneity.

The combined SNVM and MPMS study confirms that 2D $V_\alpha Cr_\beta Fe_\gamma Co_\delta Ni_{1-\alpha-\beta-\gamma-\delta}S_{2-x}$ remains magnetic with a $T_C$ well above ambient conditions. It exhibits robust intrinsic magnetism with competing spin alignments, rather than simple ferromagnetic or antiferromagnetic ordering, as evidenced by the nonmonotonic temperature evolution of $M(T)$ in Fig. 4d and the net $M(H)$ curves in Supplementary Fig. 31c. Together, these features are indicative of a magnetic ground state, in which the phase stability of the multi-element solid solution is thermodynamically governed by a substantial ideal configurational entropy ($\Delta S_{conf}$ = 1.53$R$)[39, 49]. This entropy contribution lowers the total Gibbs free energy of mixing ($\Delta G_{mix} = \Delta H_{mix} - T \cdot \Delta S_{mix}$), thereby compensating for the positive mixing enthalpy ($\Delta H_{mix}$) and suppressing phase separation into secondary binary phases. The methodology and parameters used to calculate $\Delta S_{conf}$ are provided in Supplementary Table 7 and corresponding section.

To understand the effect of configurational entropy on magnetic configurations in 2D HE-TMDs, first-principles density functional theory (DFT) simulations were carried out with a supercell structure containing 46 atoms with formula of $V_4Cr_4Fe_3Co_3Ni_4S_{28}$, where the subscripts indicate the number of atoms for each element. This is equivalent to $V_{0.22}Cr_{0.22}Fe_{0.17}Co_{0.17}Ni_{0.22}S_{1.55}$. As shown by the top and side views of the structure in Supplementary Fig. 32, the five metal elements are randomly distributed so that each metal atom's neighboring metal atom belongs to other elements. Note that bulk $VS_2$ is typically paramagnetic or weakly ferromagnetic in its 1T structural configuration[50]. Another parent binary compound $Cr_2S_3$ generally assumes a ferrimagnetic ground state where the net magnetic moment arises because the opposing antiferromagnetic sublattices do not fully cancel out due to the unequal occupancy of three Cr sites in the structure[51]. For three other parent binary compounds, including hexagonal troilite FeS, hexagonal $\alpha$-CoS, and hexagonal $\alpha$-NiS, they are often found in the A-type antiferromagnetic order, a configuration where spins are aligned ferromagnetically within the layers or planes but are aligned antiferromagnetically between the neighboring layers or planes[52, 53]. Because different binary compounds favor different magnetic configurations, it is expected that the magnetic interactions in the HE-TMDs are rich and complex due to the competing ferromagnetic and antiferromagnetic exchange interactions among different metal elements.

Therefore, in the DFT calculations, a variety of initial magnetic configurations, including ferromagnetic, A-type antiferromagnetic and intralayer antiferromagnetic orders, were considered for the HE supercell structure. After optimization of atomic coordinates, lattice vectors, and magnetic moments, four stable spin configurations were found as shown in Supplementary Table 8. Spin configuration 1 is the magnetic ground state that resembles the A-type antiferromagnetic order, but it is not exactly A-type antiferromagnetic. For instance, two Fe atoms in layer 2 assume anti-parallel spin alignment, and four Ni atoms do not follow antiferromagnetic. In fact, it has a notable non-zero total magnetic moment of 3.016 $\mu_B$ per unit cell. Spin configuration 3 is the one that is most similar to A-type antiferromagnetic, but it is less stable by 0.283 eV and it also has a non-zero magnetic moment of -4.996 $\mu_B$ per unit cell. Spin configuration 2 is close to a conventional ferromagnetic order and the second most energetically stable magnetic state. The anti-parallel spin alignments were also imposed on all metal elements within each layer (i.e., intralayer antiferromagnetic), and the subsequent relaxation gave rise to spin configuration 4 where most metal elements assume the opposite spin orientations within each layer and between the layers. It is about as stable as spin configuration 2 and has non-zero magnetization too.

The DFT results suggest that the random distributions of five metal elements in the HE-TMDs and the accompanying lattice distortions strongly influence magnetic interactions between metal cations and lead to rich and complex magnetic states that do not follow traditional ferromagnetic and antiferromagnetic orders in binary compounds. Instead, the calculations here reveal that all spin configurations in the HE-TMDs show non-zero total magnetization since the same metal element (e.g., Cr, V in Supplementary Table 8) has different atomic magnetic moments depending on its specific location and bonding configuration and thus atomic magnetic moments cannot be perfectly cancelled out even they are anti-parallel aligned. In this sense, all spin configurations in DFT calculations are ferrimagnetic, although some are closer to being antiferromagnetic while others are closer to being ferromagnetic. The ferrimagnetic ground state can explain the experimentally observed non-zero total magnetization in the samples and the fact that local magnetic moments vary from positive to negative at different locations (Supplementary Fig. 32).

The DFT calculations also show that the competing magnetic configurations differ by ~0.28 eV, more than an order of magnitude larger than kBT at room temperature. Furthermore, these represent static energy differences and the actual transition barriers are even higher, so the system cannot thermally fluctuate among these states around room temperature, and the DFT results do not support a frustrated or spin-glass-like ground state. Experimentally, NV magnetometry images coherent nanoscale magnetic domains from 300 K to 410 K, with no evidence of fragmentation or freezing, and MPMS measurements show finite coercivity and remanence up to 395 K. These behaviors are incompatible with canonical cluster-glass or spin-glass systems, which exhibit random magnetic textures, history dependence, and collapse of coercivity near the freezing temperature. The broad feature near ~320 K is therefore best interpreted as a crossover among ferrimagnetic configurations with distributed anisotropy energies rather than a spin-glass transition. Together, the DFT energetics, NV microscopy, and MPMS hysteresis consistently indicate a robust high-temperature ferrimagnetic state, not a glassy or metastable magnetic phase.

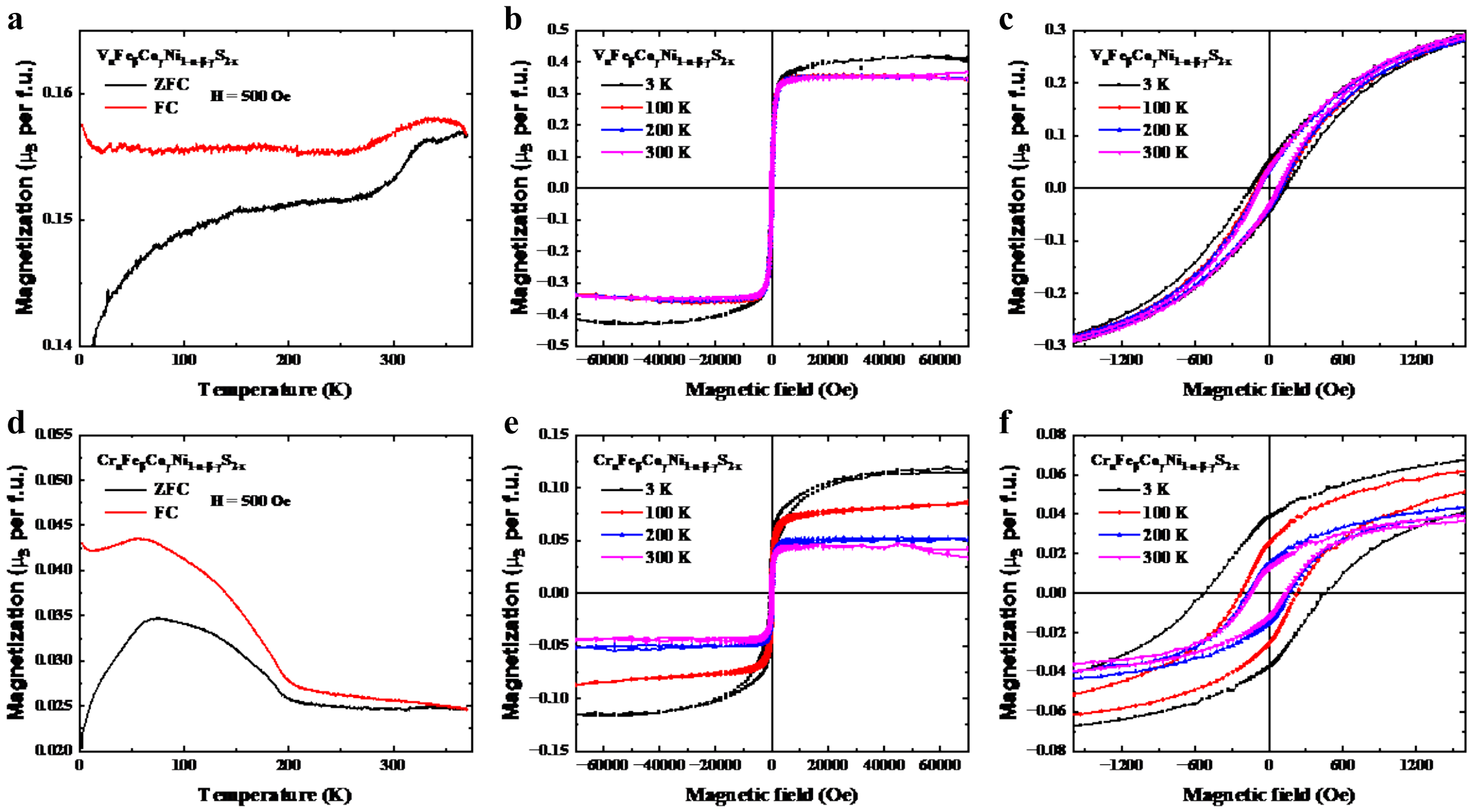

**Fig. 5| Tuning the magnetism in 2D HE-TMDs by separately excluding Cr or V.** ZFC and FC magnetization of $V_{\alpha}Fe_{\beta}Co_{\gamma}Ni_{1-\alpha-\beta-\gamma}S_{2-x}$ (**a**) and $Cr_{\alpha}Fe_{\beta}Co_{\gamma}Ni_{1-\alpha-\beta-\gamma}S_{2-x}$ (**d**). **b, e,** Magnetic hysteresis loops measured at 3, 100, 200, and 300 K, demonstrating S shape ferromagnetic hysteresis above room temperature. **c, f,** Low-field region of the *M*(*H*) loops in **b** and **e**, highlighting the persistent coercivity and remanence with increasing temperature.

Fig. 5 presents magnetic properties of four-metal 2D TMSs obtained by excluding Cr and V separately, further illustrating how elemental configuration strongly influences the competition between ferromagnetic and antiferromagnetic interactions within the system. Figs. 5a-c show temperature-dependent magnetization and field-dependent hysteresis curves for 2D $V_{\alpha}Fe_{\beta}Co_{\gamma}Ni_{1-\alpha-\beta-\gamma}S_{2-x}$, revealing a coexistence of competing ferromagnetic-like and antiferromagnetic-like features. In this composition, the antiferromagnetic contribution becomes more pronounced as the temperature decreases below ~350 K, as characterized by the downward trend in magnetization. The magnetization hysteresis exhibits a characteristic *S*-shaped curve with prominent coercivity and remanence, indicating the existence of ferromagnetic ordering despite the competing antiferromagnetic interactions. In contrast, Figs. 5d-f display typical ferromagnetic hysteresis loops for 2D $Cr_{\alpha}Fe_{\beta}Co_{\gamma}Ni_{1-\alpha-\beta-\gamma}S_{2-x}$, with well-defined coercivity and remanent magnetization persisting across all measured temperatures. The saturation magnetization increases at low temperature, and the loop shape remains symmetric and square-like, consistent with a dominant ferromagnetic phase. Compared with the previous compositions (Fig. 4, and Figs. 5 a-c), these results clearly demonstrate that tuning the constituent elements alters the relative strength of ferromagnetic and antiferromagnetic interactions.

Our DFT calculations indicate that the HE-TMDs have a ferrimagnetic ground state with non-zero total magnetization. Both V and Cr can have anti-parallel spin alignments. Therefore, substituting V for Cr does not necessarily switch off antiferromagnetic contributions. For the V compound, it is expected to be in a ferrimagnetic state with the spin up component to almost cancel the spin down component, and the total magnetization tends to increase with rising temperature since the cancellation effect is weakened due to spin disorder at higher temperatures. On the contrary, the Cr compound probably has one spin component dominant in its ferrimagnetic state because Cr element hosts significantly higher magnetic moments than V. It behaves more like a ferromagnetic state where the total magnetization decreases with increasing temperature because of enhanced spin disorder. In summary, substituting V for Cr leads to changes in the complex ferromagnetic and antiferromagnetic spin competition in the HE-TMDs, and therefore changes in the temperature dependence of the magnetic state. This comparative evolution underscores the chemical tunability of magnetism in 2D high-entropy systems where magnetic anisotropy, coercivity, and $T_C$ can all be systematically tailored through elemental substitutions. To contextualize the observed magnetic ground states, we summarized the magnetic properties of representative binary TMDs from the literature alongside our HE-TMDs in Supplementary Table 9.

## Conclusion

In summary, this work establishes a generalizable strategy for fabricating 2D HE-TMDs with intrinsic room-temperature magnetism. By combining mild hydrothermal coprecipitation of HE-LDHs with controlled chalcogenization, the compositional space of 2D materials can be expanded beyond conventional transition metal systems. This approach enables atomic-level homogeneity and structural integrity while allowing for tunable magnetic ground states from ferromagnetic to antiferromagnetic through compositional modulation. The successful translation of high-entropy design into van der Waals 2D systems provides a powerful means to tailor magnetic and electronic functionalities. Beyond magnetism, the demonstrated chemical flexibility enables tuning of correlated physical properties through control of elemental composition, potentially altering exchange fields, spin-orbit coupling, and carrier density to access antiferromagnetic, ferromagnetic, semiconducting, or topological states[54]. For example, the preliminary optical measurement on $V_{\alpha}Cr_{\beta}Fe_{\gamma}Co_{\delta}Ni_{1-\alpha-\beta-\gamma-\delta}S_{2-x}$ flakes is included in Supplementary Fig. 33.

The compositional tunability achieved here offers a promising pathway towards multifunctional spintronic devices. The ability to stabilize coexisting or switchable magnetic phases within a single material class enables reconfigurable architecture, such as spin filters, tunnel junctions, and 2D spin-logic elements. Due to their structural compatibility, these 2D systems are ideal for heterostructure integration, where interlayer coupling and proximity effects can further control spin transport and anisotropy. Future efforts should focus on composition-structure-property mapping and device-level demonstrations exploiting entropy-driven magnetic phenomena[55]. Additionally, the fabrication of monolayer HE-TMDs remains an important future direction. Combined with atomic-resolution STEM-EDS or EELS line scans characterization, such materials would enable more definitive verification of the single-phase nature, compositional homogeneity, and intrinsic properties of HE-TMDs.

## Materials and Methods

### 1. Materials

Chemicals, all used without any further purification in this study, include titanium chloride ($TiCl_4$, ≥ 99.99%), vanadium chloride ($VCl_3$, ≥99%), chromium nitrate nonahydrate ($Cr(NO_3)_3 \cdot 9H_2O$, ≥99%), manganese sulfate monohydrate ($MnSO_4 \cdot H_2O$, ≥99%), ferrous chloride tetrahydrate ($FeCl_2 \cdot 4H_2O$, ≥ 99.99%), cobalt nitrate hexahydrate ($Co(NO_3)_2 \cdot 6H_2O$, ⩾99%), nickel nitrate hexahydrate ($Ni(NO_3)_2 \cdot 6H_2O$, ⩾98%), copper nitrate hexahydrate ($Cu(NO_3)_2 \cdot 6H_2O$, ⩾99%), zirconium chloride ($ZrCl_4$, ⩾99.99%), molybdic acid ($H_2MoO_4$, ⩾85%), and tungsten chloride ($WCl_6$, ⩾99%) from Thermo Scientific Chemicals, zinc nitrate hexahydrate ($Zn(NO_3)_2 \cdot 6H_2O$, 99%), sulfur (⩾99.5%), selenium (⩾99.5%), tellurium (⩾ 99.8%), and ammonium fluoride ($NH_4F$, ⩾98%) from Alfa Aesar, acetone (⩾99.5%) and ethanol (⩾99.5%) from Fisher Chemical.

## 2. Methods

### 2.1 Synthesis of HE-LDHs

Synthesis of HE-LDHs: VCrFeCoNi HE-LDH was synthesized through a hydrothermal method. 2.5 mmol $NH_4F$, 6 mmol Urea, 1.0 mmol $Ni(NO_3)_2 \cdot 6H_2O$, 0.5 mmol $Co(NO_3)_2 \cdot 6H_2O$, 0.5 mmol $FeCl_2 \cdot 4H_2O$, 0.5 mmol $Cr(NO_3)_3 \cdot 9H_2O$, and 0.5 mmol $VCl_3$ were added to 30 ml DI water and mixed with magnetic stirring to obtain a homogeneous solution. The reaction solution was transferred into a 50 mL Teflon-lined autoclave followed by hydrothermal reaction at 150°C for 6 h. The as-synthesized VCrFeCoNi HE-LDH was washed with DI water and then dried in the oven at 60°C overnight. For the synthesis of other HE-LDHs, the steps are the same as those for VCrFeCoNi HE-LDH, except that V and/or Cr are replaced by other corresponding metal precursors.

### 2.2 Chalcogenization

Sulfurization:

The HE-LDH powders were uniformly dispersed onto a silicon wafer coated with a 300 nm $SiO_2$ layer. The wafer was then placed face-up in an alumina crucible, which was positioned at the center of a tube furnace. Another alumina crucible containing 500 mg of sulfur powder was placed upstream along the gas flow direction, and its distance from the furnace center was adjusted to maintain the sulfur source at approximately 400 °C during the reaction. Prior to heating, the system was purged with Ar (500 sccm) for 30 min to remove residual air. Subsequently, the carrier gas was switched to a mixture of Ar (80 sccm) and $H_2$ (20 sccm), and the furnace temperature was ramped to the target sulfurization temperature depending on the HE-LDH composition: ZrVFeCoNi (750 °C), TiVFeCoNi (750 °C), VCrFeCoNi (850 °C), VWFeCoNi (800 °C), VCoNiCuZn (700 °C), and MnFeCoNiZn (850 °C). After reaching the set temperature, the reaction was maintained for 30 min, followed by rapid cooling to room temperature.

Selenization:

The selenization process followed the same procedure as described for sulfurization, except that the alumina crucible containing selenium powder was positioned at a location corresponding to approximately 500 °C along the upstream region of the furnace.

Tellurization:

The tellurization process was conducted following the same procedure as sulfurization, with the only difference being that the alumina crucible containing tellurium powder was placed at a position corresponding to approximately 700 °C in the upstream region of the furnace.

The whole chalcogenization process was carried out under atmospheric pressure.

### 2.3 Characterization

Morphological and structural analyses:

The microscale morphology of HE-LDHs and HE-TMDs was investigated using a FEI Quanta-600F field-emission environmental SEM, which was equipped with a Bruker Xflash 6 SDD EDS detector for elemental identification and spatial distribution analysis. For nanoscale characterization, STEM and EDS characterization were performed on the probe-corrected JEOL JEM-ARM200F NEOARM at 200 kV with 28 mrad convergence angle and 68-280 mrad HAADF collection angle with a dual large solid angle JEOL

EDS detector at Oak Ridge National Laboratory (ORNL). High-resolution TEM observations were carried out on a Thermo Fisher Scientific Titan microscope operated at 300 kV at ORNL. The surface morphology of the flakes was characterized using a Cypher S atomic force microscope (AFM, Asylum Research, Oxford Instruments) operating in tapping mode under ambient conditions. Silicon cantilevers (Titan 70, Oxford Instruments; nominal resonance frequency ≈ 70 kHz, spring constant ≈ 2 $N \cdot m^{-1}$, aluminum-coated backside) were excited near their mechanical resonance. The resonance frequency was identified through an auto-tuning sweep prior to image acquisition. During measurements, the oscillation amplitude setpoint was adjusted to 70-80% of the free amplitude to ensure gentle tip-sample interaction and stable tapping performance. The crystallinity and phase composition of HE-LDHs and HE-TMDs were examined using a Rigaku Smart Lab X-ray diffractometer (XRD) with Cu Kα radiation.

Chemical state measurements:

X-ray photoelectron spectroscopy (XPS) measurements were performed on a Quantera XPS system using an Al Kα X-ray source. Raman spectroscopy was carried out using a Renishaw inVia$^{TM}$ Raman microscope with a 633 nm laser to investigate the vibrational modes of the samples.

Magnetic behavior investigation:

NV magnetometry measurements were carried out using a commercial room-temperature scanning NV microscope (ProteusQ™, Qnami). The probe consisted of a diamond cantilever with an integrated nanopillar hosting a single NV center, positioned ~50 nm above the sample surface. The NV tip has ~200 nm apex diameter, which is larger than a conventional AFM tip. This larger tip can facilitate enhanced photon collection efficiency but at the expense of lateral resolution. The NV was optically excited with a 520 nm laser (~20 μW), and the resulting photoluminescence was collected through a confocal optical setup. Microwave control (~0.1 W input) was provided by a shorted coaxial antenna located ~15 μm from the NV and driven by a Qnami MicrowaveQ signal generator. A permanent magnet (56 G) oriented along the surface normal was applied to the NV center and the sample during the measurement, and its contribution was subtracted from the recorded stray field maps.

Magnetic measurements were performed using a Quantum Design Magnetic Property Measurement System (MPMS) at the Center for Nanophase Materials Science (CNMS) at Oak Ridge National Laboratory. The field cooled (FC) and zero field cooled (ZFC) measurements were carried out between the temperature range of 2-400 K. Magnetic hysteresis measurements were taken with a maximum sweeping magnetic field of 70000 Oe on the same instrument at 3 K, 100 K, 200 K and 300 K.

**3. DFT Calculations**

First-principles DFT calculations were performed using the VASP package, where projector augmented wave (PAW) pseudopotentials were used for electron-ion interactions and the Perdew-Burke-Ernzerhof (PBE) functional was used for exchange-correlation interactions[56, 57]. Van der Waals (vdW) interactions were included using the DFT-D3 method. The cutoff energy was set at 350 eV. To capture the strong correlation effects from localized d orbitals, we adopted the DFT+U method using Dudarev's rotationally invariant approach with an effective Hubbard parameter of 2.0 eV for V and 3.0 eV for other metal elements[58-60]. For a bulk unit cell of $VS_2$ in the 1T phase (only one layer), both atomic positions and lattice constants were allowed to relax until the residual forces below 0.01 eV/Å, with a 24×24×12 k-point sampling in the Brillouin zone. Based on the optimized unit cell, a 3×3×2 supercell was built with the k-point sampling reduced to 8×8×6 accordingly. In this supercell, it contains two layers for us to simulate different interlayer magnetic orders. To simulate experimental 2D HE TMDs structures, we then randomly substituted V atoms with Cr, Fe, Co, and Ni atoms in the supercell and removed 8 S atoms as well, resulting in a $V_4Cr_4Fe_3Co_3Ni_4S_{28}$ where the subscripts indicate the number of atoms for each element. A variety of initial magnetic configurations were imposed on the supercell structure for subsequent structural relaxation and spin optimization. For this HE supercell structure, both atomic positions and lattice constants were optimized until the residual forces below 0.01 eV/Å, which allows lattice distortions and atomic reconstructions to occur to mimic the realistic condition.

**Acknowledgments**

Y.Y. gratefully acknowledges financial support by the U.S. National Science Foundation, Faculty Early Career Development Program (CAREER), under award number 2420622. SNVM, MPMS, TEM, and DFT research conducted as part of a user project at the Center for Nanophase Materials Sciences (CNMS), which is a US Department of Energy, Office of Science User Facility at Oak Ridge National Laboratory. Y.Y., C.Z. and X.Z. also acknowledge the characterization support from the CNMS at ORNL under user proposals CNMS2023-B-02166 and CNMS2025-B-03380. L.L. acknowledges computational resources of the Compute and Data Environment for Science (CADES) at the Oak Ridge National Laboratory, which is supported by the Office of Science of the U.S. Department of Energy under Contract No. DE-AC05-00OR22725.

**Author Contributions**

Y.Y., Z.G., and H.Z. conceived the project. C.Z., X.Z. and B.G.G. designed and carried out the fabrication experiments. W.W. carried out XPS analysis. C.T.N., X.H. and H.C. conducted TEM imaging. M.C. and X.W. helped with AFM measurement. D.K. and Z.G. carried out MPMS analysis. H.Z. and A.L.M. performed NV testing. P.L. and X.Y. carried out electrical measurements. L.L. conducted DFT calculations. C.Z. and X.Z. supervised by Y.Y., Z.G. and H.Z. analyzed all data and drafted the manuscript with input, discussion and approval from all co-authors.

**Data Availability**

The data generated in this study are provided in the Source Data file accompanying this paper. All other raw data relevant to this work are available from the corresponding author upon request.

**Competing Interests**

The authors declare no competing interests.

**Supplementary Information**

This file contains Supplementary Information, which includes Supplementary Figs 1-33, Supplementary Tables 1-9, Supplementary Texts, and Supplementary References.